\documentclass[11pt,a4paper]{article}
\usepackage{graphicx}
\usepackage{amsmath}
\usepackage{cite}
\usepackage{verbatim} 
\usepackage[usenames]{color}

\newcommand{\bEq}{\begin{equation}}
\newcommand{\eEq}{\end{equation}}
\newcommand{\bEQ}[1]{\begin{equation} \begin{array}{#1}}
\newcommand{\eEQ}{\end{array} \end{equation}}

\newcommand{\correction}[1]{\textcolor{black}{#1}}

\begin{document}

\pagestyle{empty}

\null

\vfill

\begin{center}

{\Large  
{\bf Indexing neutron transmission spectra of rotating crystal}}\\

\vskip 1.0cm
A. Morawiec
\vskip 0.5cm 
{Institute of Metallurgy and Materials Science, 
Polish Academy of Sciences, \\ Krak{\'o}w, Poland.
}
\\
E-mail: nmmorawi@cyf-kr.edu.pl \\
Tel.: ++48--122952854, \ \ \  Fax: ++48--122952804 \\

 \end{center}

\vfill

\noindent
{\bf Synopsis}\\
Neutron transmission spectra of a rotating crystal carry information about its lattice.
The relationship between Bragg-dip positions in the spectra and the crystal lattice 
is investigated theoretically, and
schemes for unambiguous determination of crystal orientation and lattice parameters 
are presented. 


\vfill

\noindent
{\bf Abstract}
\\
Neutron \correction{time-of-flight} transmission spectra of mosaic crystals 
contain Bragg dips, i.e., 
minima at wavelengths corresponding to diffraction reflections.
Positions of the dips 
are used for investigating crystal lattices.
By rotating the sample around a fixed axis 
and recording spectrum at each rotation step, 
the intensity of the transmitted beam is obtained 
as a function of the rotation angle and wavelength.
The questions addressed concern determination of lattices parameters and orientations of
centrosymmetric crystals from such data. 
It is shown that if the axis of sample rotation is inclined to the beam direction, 
the reflection positions unambiguously determine reciprocal lattice vectors, 
which is not the case when the axis is perpendicular to the beam.
Having a set of such vectors, one can compute crystal 
orientation or lattice parameters using existing indexing software.
The considerations are applicable to arbitrary Laue symmetry.
The work contributes to the automation of analysis of 
diffraction data obtained in the neutron imaging mode.

\vskip 0.5cm


\vskip 0.2cm

\noindent
\hfill \today

\newpage

\pagestyle{plain}

\noindent
\section{Introduction}

Neutron studies of crystalline materials rely mainly on scattering methods,
but neutron imaging can also be useful in research on such materials.
In simple terms, neutron imaging involves measuring the attenuation 
of the transmitted beam and usually means radiography or tomography \cite{Anderson_2009}.
Of interest here is the wavelength-resolved imaging \cite{Santisteban_2001,Woracek_2018}.
Neutron \correction{spectra recorded
in transmission imaging geometry using the time-of-flight technique}
carry information about crystal structure.
Intensity 
of the transmitted beam is affected by 
crystal diffraction, and analysis of the recorded intensities allows for drawing conclusions 
about the geometry of crystal lattice.
With this approach, crystallographic information is extracted from 
diffraction effects observed using a single wavelength-resolving point detector, 
and data are collected on imaging beamlines. 
The spatial resolution of wavelength-resolved transmission imaging is 
significantly better than that of conventional neutron diffraction \cite{Woracek_2018}.

Neutron transmission spectra of mosaic crystals contain
minima (known as Bragg dips) at locations corresponding to diffraction reflections 
\cite{Halpern_1941,Frikkee_1975,Thiyagarajan_1998,Santisteban_2005}.
Positions and shapes of the
dips depend on material, orientation  and the degree of perfection of the crystal.
They can be used to determine the beam direction in the crystal reference frame 
and to study  
crystal orientation, mosaicity, lattice parameters or elastic strain \cite{Malamud_2016}. 
Individual 'Bragg-dip patterns' have been used for 
'orientation'  and strain mappings 
\cite{Sato_2017,Sato_2018b,Strickland_2020,Watanabe_2020,Sakurai_2021,Shishido_2023,Watanabe_2024}.
However, the crystallographic information contained in a single spectrum is insufficient 
to fully characterize crystal orientation or strain.
The information is richer when multiple spectra are collected at various sample orientations.
The simplest way to vary these orientations is to rotate the sample around a fixed axis in equiangular steps.
By recording a spectrum at each rotation step, 
one obtains 
the intensity of the transmitted beam as
a function of the rotation angle $\varphi$ and the wavelength $\lambda$
\cite{Cereser_2017,Dessieux_2023a,Dessieux_2023b}.
The function can be seen as a diffraction pattern in $(\varphi,\lambda)$ coordinates.

\correction{Combining wavelength--resolved imaging with detection of diffraction spots 
enabled  Cereser et al. \cite{Cereser_2017} 
to map orientations of a multigrain sample in three dimensions.}
That mapping required a fully automatic procedure.
It was based on forward 'indexing', i.e., orientations were determined
by comparing experimental patterns with patterns 
simulated for all possible orientations.
Forward 'indexing' avoids the use of a 'backward' 
relationship leading from pattern geometry to crystal orientation.
However, there are advantages to knowing and understanding this relationship. 
It can be used to search for experimental configurations 
favorable for determination of the lattice or its orientation.

The questions addressed here concern characterization of the crystal lattice
using an experimental setup in which the sample rotation axis 
is not necessarily perpendicular to the beam direction.
Having multiple spectra collected from a centrosymmetric crystal rotated 
about a fixed axis, what are the algorithms for determining crystal orientations
or methods for estimating lattice parameters?
Closely related are the issues of uniqueness of the resulting orientations 
and parameters.

It is shown below that if the axis of sample rotation is inclined to the beam direction, 
reciprocal lattice vectors can be unambiguously determined from positions of Bragg dips in 
neutron transmission spectra. 
(\correction{There is an ambiguity in the determination 
of these vectors when the axis 
is perpendicular or parallel to the beam.})
Knowledge of several vectors of a lattice is the foundation for its full characterization.
Schemes for both orientation determination and ab initio indexing  
are described. They are based on existing software. 
The presented methods are applicable to arbitrary Laue symmetry.

Throughout the article, vectors are identified with one-column matrices. 
Unit vectors are denoted by the `hat' symbol \ $\widehat{\null}$\,.
Rotation by the angle $\alpha$ about axis parallel to the vector $\widehat{\mathbf{n}}$ 
will be denoted by $R(\widehat{\mathbf{n}},\alpha)$.
The same symbol will be used for the $3 \times 3$ special orthogonal matrix representing the rotation. 
The matrix is defined in such a way that the result of its operation on vector $\mathbf{v}$ is 
$R(\widehat{\mathbf{n}},\alpha) \mathbf{v} = 
\cos \alpha \, \mathbf{v} + 
(1-\cos \alpha) (\widehat{\mathbf{n}} \cdot \mathbf{v}) \, \widehat{\mathbf{n}}
+ \sin \alpha \,  \mathbf{v}  \times \widehat{\mathbf{n}}
$.

\section{Pattern of sinusoidal curves}

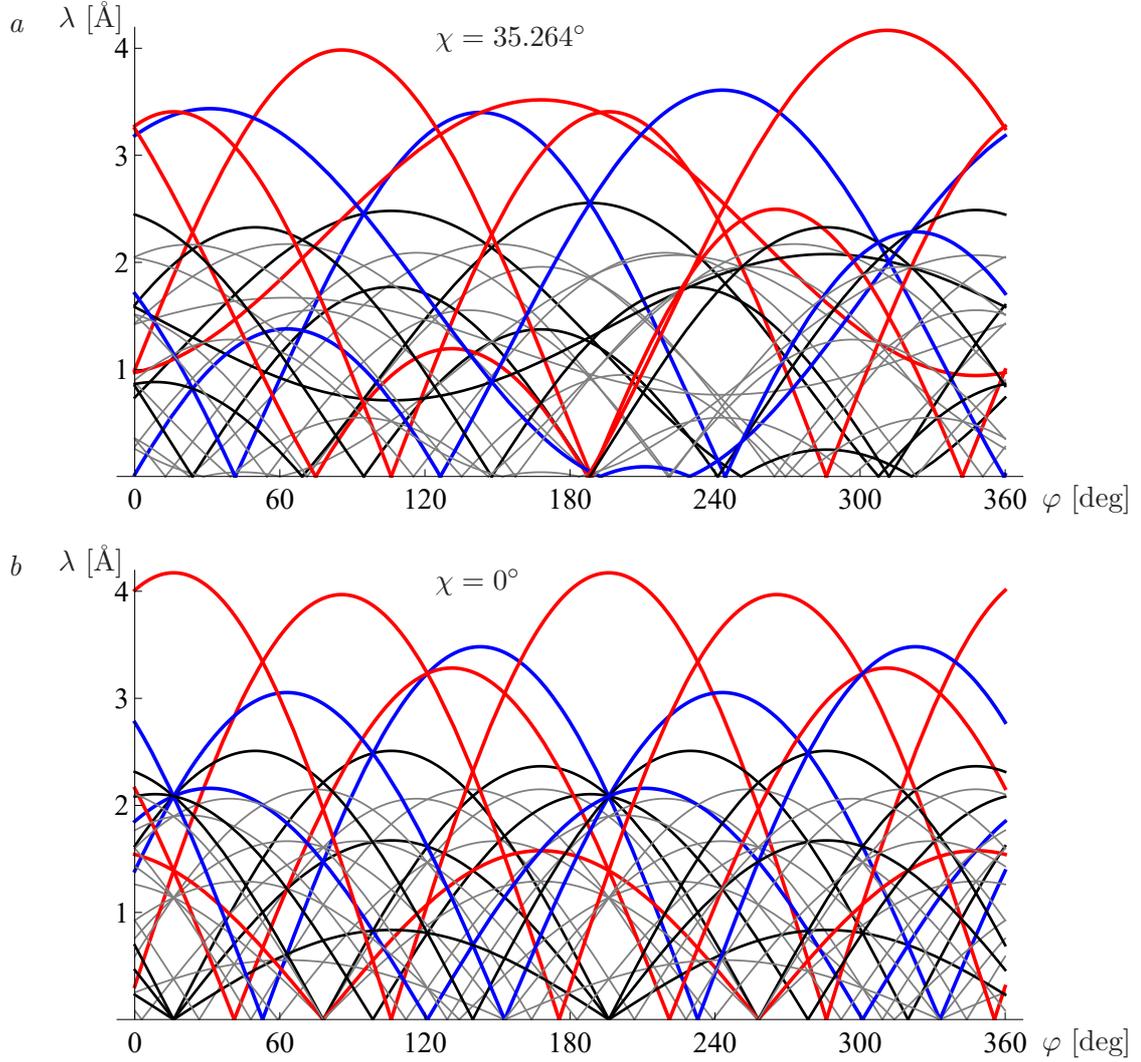
\begin{figure}
	\begin{picture}(300,390)(0,0)
		\put(50,205){\resizebox{12.0 cm}{!}{\includegraphics{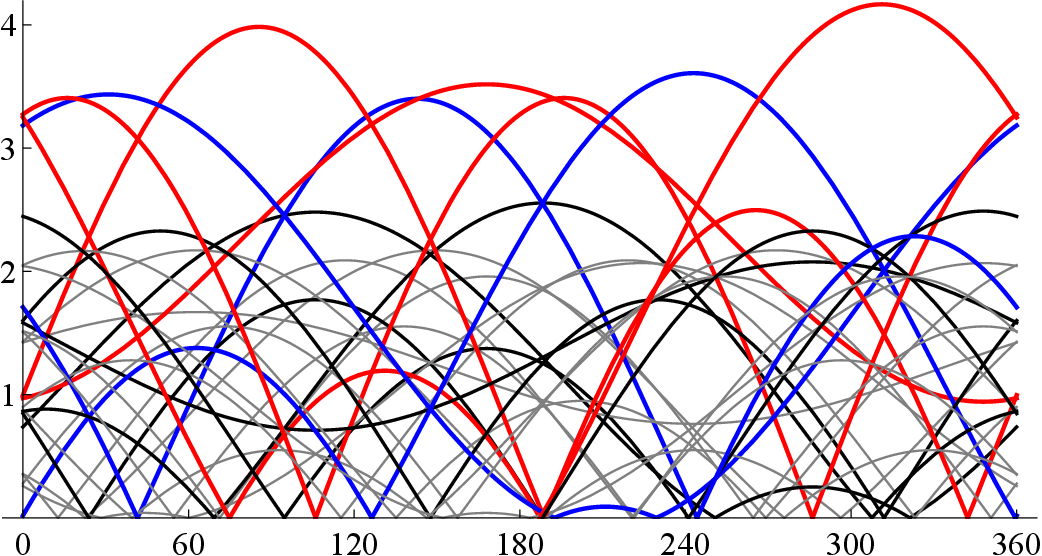}}}
		\put(50,0){\resizebox{12.0 cm}{!}{\includegraphics{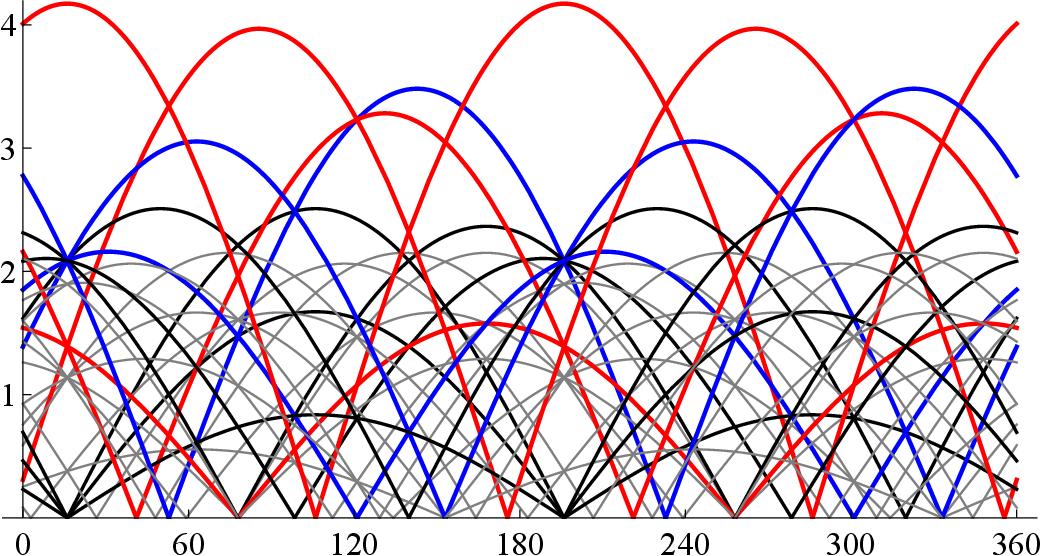}}}
		\put(10,385){\textit{a}}
		\put(29,387){$\lambda$ [\AA] }
		\put(397,206){$\varphi$ [deg] }
		\put(170,380){$\chi=35.264^{\circ}$ }
		\put(10,180){\textit{b}}
		\put(29,182){$\lambda$ [\AA] }
        \put(397,1){$\varphi$ [deg] }
		\put(170,175){$\chi=0^{\circ}$ }
	\end{picture}
	\vskip 0.0cm
	\caption{Simulated patterns of sinusoidal curves for Cu crystal 
with the plane  $(1\,2\,3)$ perpendicular to the sample rotation axis $\mathbf{e}^{\tt{S}}_3$
and the crystal direction $[\overline{6}\,\overline{3}\,4]$ along sample's $\mathbf{e}^{\tt{S}}_1$ direction.
(The orientation can be seen as a variant of the orientation $S$
common in textures of fcc metals.) 
		The lattice parameter $a=3.61334$\mbox{\AA} was used.
		Families of reflecting planes are $\{111\}$ (red), $\{200\}$ (blue), $\{220\}$ (black), $\{311\}$ (gray).
		The angle $\chi$ between the plane perpendicular to the incident beam 
		and the axis of specimen rotation is 
		$35.264^{\circ}$ in (\textit{a}) and $0^{\circ}$ in (\textit{b}).
Periods of the patterns in (\textit{a}) and (\textit{b}) are $360^{\circ}$ and $180^{\circ}$,
respectively.
	}
	\label{Fig_completePatternS}
\end{figure}

A Bragg dip in the spectrum is the result of crystal diffraction.
First, one needs to link the wavelength 
at which the dip occurs with the reciprocal lattice node corresponding to the dip.
Let $\mathbf{k}_0$ and $\mathbf{k}$ be the wavevectors of the incident and diffracted beams,
respectively.
The geometry of crystal diffraction is described by 
the energy conservation law $| \mathbf{k} | = 1/\lambda = |\mathbf{k}_0 |$
and the property that the scattering vector $\mathbf{k}-\mathbf{k}_0$
points to a node of the crystal reciprocal lattice.
Analysis of crystal diffraction data 
involves three right-handed Cartesian reference systems: 
the one associated with the laboratory (\correction{indicated by the superscript $\tt{L}$}),
the sample reference system ($\tt{S}$), and the system attached to the crystal ($\tt{C}$). 
Vectors with components in different reference frames
will be denoted by different symbols. 
Let $\mathbf{g}^{\tt{L}}$ be the array with components of 
a reciprocal lattice vector in the laboratory reference frame.
With the wavevectors given in the same reference frame, 
based on $\mathbf{k}-\mathbf{k}_0 = \mathbf{g}^{\tt{L}}$, 
one has $(\mathbf{k}+\mathbf{k}_0) \cdot \mathbf{g}^{\tt{L}} = 
\mathbf{k}^2 -\mathbf{k}_0^2=0$,
and elimination of $\mathbf{k}$ leads to 
\bEq
\mathbf{g}^{\tt{L}} \cdot (\mathbf{g}^{\tt{L}}+2 \mathbf{k}_0) = 0 \ .
\label{eq:Laue2}
\eEq
Let 
the reciprocal lattice vector 
in the Cartesian reference frame attached to the crystal
be denoted by $\mathbf{h}$.\footnote{For notational 
consistency, 
this vector could be denoted as $\mathbf{g}^{\tt{C}}$, but the symbol $\mathbf{h}$ is preferred 
to have the same notation 
as in related papers referred to below.
The same applies to $\mathbf{g}=\mathbf{g}^{\tt{S}}$.}
With $O$ standing for the special orthogonal matrix 
representing crystal orientation in the sample reference frame 
(Bunge's convention \cite{Bunge_1982,Morawiec_2004}), 
the array $\mathbf{g}=O^T \mathbf{h}$ has components in the sample reference frame
and is related to $\mathbf{g}^{\tt{L}}$ via 
\bEq
\mathbf{g}^{\tt{L}} = R_{\varphi} \,\mathbf{g} 
\ , 
\label{eq:g_h}
\eEq
where 
$R_{\varphi}$ is the special orthogonal matrix representing the ($\varphi$-dependent) orientation of the sample reference frame
with respect to the laboratory reference frame.
Substitution of $\mathbf{g}^{\tt{L}}$ given by (\ref{eq:g_h}) into (\ref{eq:Laue2}) leads to 
\bEq
\mathbf{h}^2  +2 ( R_{\varphi} \, O^T \mathbf{h} ) \cdot \mathbf{k}_0 = 0 \ ,
\label{eq:intemediate}
\eEq
where $\mathbf{h}^2$ is an abbreviation for $\mathbf{h} \cdot \mathbf{h}$.
With $\widehat{\mathbf{k}}_0 = \lambda \mathbf{k}_0$, one has
$\lambda(\varphi)   = -(2 /\mathbf{h}^2) \ ( O^T \mathbf{h} ) \cdot (R_{\varphi}^T \widehat{\mathbf{k}}_0)$,
or briefly
\bEq
\lambda(\varphi)   = \widehat{\mathbf{k}}^{\tt{S}}(\varphi) \cdot  \mathbf{d} \ , 
\label{eq:for_g}
\eEq
where 
$
\widehat{\mathbf{k}}^{\tt{S}}(\varphi) = R_{\varphi}^T \widehat{\mathbf{k}}_0 
$
and
$
\mathbf{d} =  -(2 /\mathbf{h}^2) \, O^T  \mathbf{h} = - (2 /\mathbf{g}^2) \,  \mathbf{g}
$.
The wavelength $\lambda$ as a  function of $\varphi$  is a sinusoid of period $2 \pi$.
Only parts of the sinusoid with $\lambda(\varphi) > 0$ are physically meaningful.
Since  multiple reflecting planes (and corresponding vectors $\mathbf{h}$) are involved, 
multiple sinusoidal curves appear in a pattern.
Example patterns of sinusoids in 
$(\varphi,\lambda)$ coordinates are shown in Fig.~\ref{Fig_completePatternS}. 

Knowing $\mathbf{d}$,  one can calculate the corresponding reciprocal lattice vector 
in the sample reference frame
\bEq
\mathbf{g} =  -(2 /\mathbf{d}^2) \,  \mathbf{d} \  .
\label{eq:for_gs}
\eEq
Therefore, the prior step is to determine the sinusoid parameters $\mathbf{d}$ 
from  $\lambda(\varphi)$  and $\widehat{\mathbf{k}}^{\tt{S}}(\varphi)$ 
using eq.(\ref{eq:for_g}).

\section{Parameters of sinusoids}

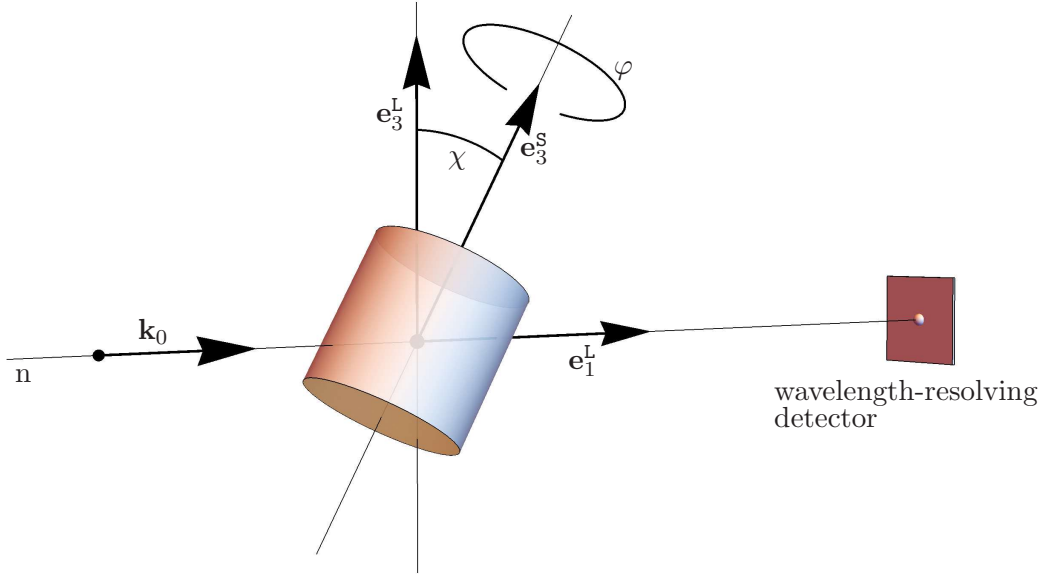
\begin{figure}
	\begin{picture}(300,220)(0,0)
		\put(0,-30){\resizebox{14.0 cm}{!}{\includegraphics{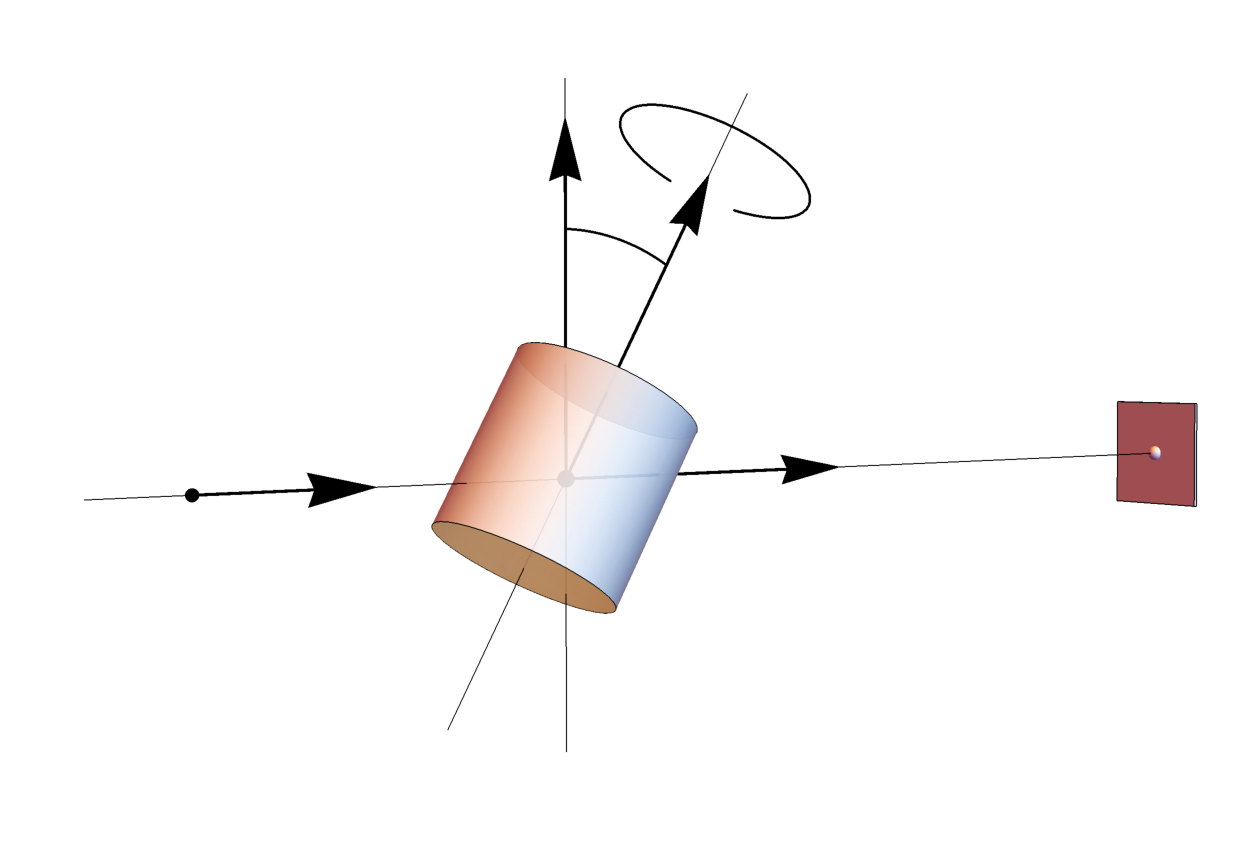}}}	
		\put(30,77){n}
		\put(314,70){wavelength-resolving}
		\put(314,60){detector}
		\put(76,93){$\mathbf{k}_0$}
		\put(219,162){$\mathbf{e}^{\tt{S}}_3$}
		\put(165,175){$\mathbf{e}^{\tt{L}}_3$}
		\put(237,82){$\mathbf{e}^{\tt{L}}_1$}
		\put(192,159){$\chi$}
		\put(254,193){$\varphi$}
	\end{picture}
	\vskip 0.0cm
	\caption{Schematic geometry of experimental setup with sample rotation axis 
	inclined to the neutron beam direction.}
	\label{Fig_Bragg_dip}
\end{figure}

Without loosing generality, one can assume that the basis vector 
$\mathbf{e}^{\tt{L}}_1$ of the laboratory reference frame
is along $\mathbf{k}_0$, 
$\mathbf{e}^{\tt{S}}_3$ is along the axis of crystal rotation and lies 
in the plane spanned by $\mathbf{k}_0$ and $\mathbf{e}^{\tt{L}}_3$,
and the scalar products 
$\mathbf{k}_0 \cdot \mathbf{e}^{\tt{L}}_1$,
$\mathbf{k}_0 \cdot \mathbf{e}^{\tt{S}}_3$ and $\mathbf{e}^{\tt{L}}_3 \cdot \mathbf{e}^{\tt{S}}_3$
are non-negative (Fig.~\ref{Fig_Bragg_dip}).
With these assumptions and fixed $\chi=\arccos(\mathbf{e}^{\tt{L}}_3 \cdot \mathbf{e}^{\tt{S}}_3)$, 
one has $\widehat{\mathbf{k}}_0 =[1\ 0\ 0]^T$,
$0 \leq \chi \leq \pi/2$,
$R_{\varphi} = R(\mathbf{e}^{\tt{L}}_2,\chi)^T R(\mathbf{e}^{\tt{S}}_3,\varphi)$ and
$\widehat{\mathbf{k}}^{\tt{S}}(\varphi) = \left[\cos\chi \, \cos \varphi \ \ \cos \chi  \sin \varphi \ \ 
\sin \chi \right]^T$.

Knowing wavelengths on a sinusoid at three different angles 
$\varphi_j$ 
($j=1,2,3$ and $0 \leq \varphi_j < 2 \pi$), 
one has a system of three linear equations
\bEq
\lambda(\varphi_j)  = \widehat{\mathbf{k}}^{\tt{S}}(\varphi_j)  \cdot \mathbf{d}  
\label{eq:system}
\eEq
for components of $\mathbf{d}$.
The coefficient matrix is the transpose of 
$K_{\tt{S}}=
[\widehat{\mathbf{k}}^{\tt{S}}(\varphi_1) \ \widehat{\mathbf{k}}^{\tt{S}}(\varphi_2) 
\ \widehat{\mathbf{k}}^{\tt{S}}(\varphi_3)]$. 
Its determinant is given by
\bEq
\det(K_{\tt{S}})=\sin \chi \cos^2 \chi \  
(\sin (\varphi_1-\varphi_3)+\sin (\varphi_2-\varphi_1)+\sin (\varphi_3-\varphi_2)) \ .
\label{eq:determinant}
\eEq
The determinant equals zero 
when $\chi$ equals $0$ or $\pi/2$.
Clearly, there is an essential difference between singular 
and non-singular 
cases; 
if the matrix $K_{\tt{S}}$ is singular, $\mathbf{d}$ is not unique, and 
this affects the determinability of the reciprocal lattice vector $\mathbf{g}$.

\subsection{Singular cases } 

\noindent
\textit{The case of $\chi=\pi/2$}
\\
The physical reason for the singularity at $\chi=\pi/2$
is obvious: Rotation of the sample  about axis parallel to the beam
does not affect the angles between reflecting planes and the beam. 
In this case, the sinusoids become straight lines, and since 
$\widehat{\mathbf{k}}^{\tt{S}}(\varphi)=[0 \ 0 \ 1 ]^T$, the wavelength is determined by the 
third component of $\mathbf{d}$.
Clearly, spectra recorded during sample rotation 
contain the same information as a single spectrum.
Indexing an individual spectrum
is equivalent to determination of the transmitted beam direction
in the crystal reference frame.  
This subject is addressed in Appendix. 
It is easy to see that the orientations
$O R(\mathbf{e}^{\tt{S}}_3,\xi)$ lead to the same pattern regardless of the angle $\xi$. 
Moreover, 
the same pattern is obtained after
the sample is half-turned about an axis perpendicular to the beam.
Thus, also the orientations 
$O R(\mathbf{e}^{\tt{S}}_1,\pi) R(\mathbf{e}^{\tt{S}}_3,\xi)$
lead to the same pattern as that for $O$.

\vskip 0.2cm

\noindent
\textit{The case of $\chi=0$}
\\
The reason for the singularity at $\chi=0$ is more subtle: 
With varying $\varphi$, the impact of 
the change in the angle between the beam and the reflecting plane 
on the last component of $\mathbf{d}$
is canceled out by the change in the interplanar spacing. 
Since $\widehat{\mathbf{k}}^{\tt{S}}(\varphi)=[\cos \varphi \ \sin \varphi \ 0 ]^T$, 
the third component of $\mathbf{d}$ does not affect the sinusoid $\lambda(\varphi)$.

With $\chi=0$, the pattern of sinusoids
has the period of $\pi$ along $\varphi$, 
i.e., the spectra at $\varphi$ and $\varphi+\pi$ are identical:
Since $\mathbf{h}^2=(-\mathbf{h})^2$ and 
$
( R_{\varphi} \, O^T \mathbf{h} ) \cdot \mathbf{k}_0 = 
(- O^T \mathbf{h} ) \cdot ( -R_{\varphi}^T \,\mathbf{k}_0)=
{( O^T (-\mathbf{h}) ) \cdot ( R_{\varphi+\pi}^T \,\mathbf{k}_0)} =
( R_{\varphi+\pi} O^T (-\mathbf{h}) ) \cdot  \mathbf{k}_0 
$,
if eq.(\ref{eq:intemediate}) is satisfied, so is the relationship
${(-\mathbf{h})^2  +2 ( R_{\varphi+\pi} O^T (-\mathbf{h}) ) \cdot \mathbf{k}_0} = 0$.
Thus, if  there is a dip at $(\varphi,\lambda)$ 
due to  $\mathbf{h}$, then there is a dip at $(\varphi+\pi,\lambda)$ due to $-\mathbf{h}$.
The fact that the pattern of sinusoids has the period of $\pi$ affects orientation determination. 
Let $C$ denote the half-turn about $\mathbf{e}^{\tt{S}}_3$, i.e., 
$C= R(\mathbf{e}^{\tt{S}}_3,\pi) =R_{\pi}$. 
Since $R_{\varphi+\pi} O^T=R_{\varphi} C O^T = R_{\varphi}  (OC)^T$, 
the periodicity implies that 
the patterns for $O$ and $O C$ are identical.\footnote{The problem of orientation ambiguity is mentioned in \cite{Cereser_2017}
	and it is extensively addressed in Supplementary information to that paper.
	It is claimed that the orientations 
	$O$ (which equals $U^T$ in notation of \cite{Cereser_2017}) and $C O C$ lead to the same pattern. 
	Generally, this claim is false, i.e., patterns from crystals at orientations  
	$O$ and $C O C$ are different.
	In the case of cubic crystals with the Cartesian axes along the four-fold symmetry axes,
	the presence of $C$ on the left side of $O$ is inconsequential  because the orientations $O$ and $CO$ are equivalent due to the crystal symmetry, so also the patterns for $OC$ and $C O C$ are identical. 
} 
This means that with $\chi=0$, there is an ambiguity in determining crystal orientation:
two equally correct orientations $O$ and $O C$ result from each pattern.
A scheme for determining the orientations 
$O$ and $O C$ based on data obtained with the crystal rotation axis 
perpendicular to the beam direction is sketched at the end of Appendix.

Summarizing,  the usual measurement geometry (see, e.g., \cite{Santisteban_2005,Malamud_2016,Cereser_2017,Dessieux_2023a})
with $\chi=0$ 
is not optimal for lattice characterization. 
There are non-equivalent crystal orientations leading to identical patterns. 
Since this geometry does not allow for 
calculating the last 
component of $\mathbf{d}$,
the reciprocal lattice vectors are not fully determinable,
which makes the geometry inconvenient for calculating lattice parameters or strain tensors.

\subsection{The case with inclined rotation axis \label{sec:incl}}

All components of $\mathbf{d}$ can be obtained when the absolute 
value of the determinant (\ref{eq:determinant}) 
is sufficiently large.
It reaches maximum when 
$\chi= \arctan \left(1/\sqrt{2}\right) \approx 35.264^{\circ}$ 
and  the separation between
the angles $\varphi_i$ is the largest possible.

The simplest scheme for computing $\mathbf{d}$ is to use $\varphi$ and $\lambda$ 
of appropriately selected points on a sinusoid. 
With $\chi$ significantly different from $0$ and $\pi/2$,  
three such points 
are sufficient, 
but clearly, a better approach is to use more, say $J$, points 
and compute $\mathbf{d}$ by solving  the linear least-squares problem 
$\min \sum_{j=1}^J ( \widehat{\mathbf{k}}^{\tt{S}}(\varphi_j)  \cdot \mathbf{d}-\lambda(\varphi_j) )^2$.
Formally, its solution is
\bEq 
\mathbf{d}= \left(K_{\tt{S}}^+ \right)^T   \boldsymbol{\lambda} \ ,
\label{eq:fin_d}
\eEq 
where 
$\boldsymbol{\lambda}=[ \lambda(\varphi_1)\, \lambda(\varphi_2)\, \ldots \, \lambda(\varphi_J)]^T$,
$K_{\tt{S}}=[ \widehat{\mathbf{k}}^{\tt{S}}(\varphi_1) \, \widehat{\mathbf{k}}^{\tt{S}}(\varphi_2) \, \ldots \, \widehat{\mathbf{k}}^{\tt{S}}(\varphi_J)]$, 
and $K_{\tt{S}}^+$ is the pseudoinverse of $K_{\tt{S}}$.
In practice, pseudoinverse is computed using singular value decomposition.

\correction{If the axis inclination angle $\chi$ is affected 
	by a small non-zero error $\varepsilon$  (in radians),
	then the relative error of the components $d_1$ and $d_2$ of $\mathbf{d}$
	is $\Delta_{1,2} \approx \varepsilon \tan \chi$ and 
	the relative error of $d_3$ is $\Delta_3 \approx \varepsilon \cot \chi$.
	(Consistent with what was written in the previous section, 
	$\Delta_{1,2}$ and $\Delta_3$ become infinite when $\chi=\pi/2$ and $\chi=0$, respectively.)
	Analogous expressions for errors of components of $\mathbf{g}$
	involve $\lambda(\varphi_j)$ and are complicated,
	but with $\chi$ near $\pi/4$, 
	the relative errors of components of $\mathbf{d}$ are 
	close to $\varepsilon$, and 
	if all components of $\mathbf{d}$ are modified with the same 
	relative error $\varepsilon$, then the relative error of 
	components of $\mathbf{g} = - 2 \mathbf{d}/\mathbf{d}^2$ is 
	$\varepsilon/(1+\varepsilon) \approx \varepsilon$. 
This gives a rough estimate of the sensitivity of 
the reciprocal lattice vector determination to the inclination angle $\chi$.
	}

\subsection{Detection of sinusoids by image analysis}

When intensities of the transmitted neutron beam are recorded 
for the angles $\varphi_j$ changed in equal steps and discrete 
equidistant wavelengths $\lambda_k$,
one obtains a dataset of intensities numerated by $j$ and $k$.
The set can be seen as a pixelated gray-scale image from which 
the parameters $\mathbf{d}$ of individual sinusoids are to be extracted.
This can be done automatically. 
The subject of 
automatic determination of $\mathbf{d}$ by image analysis techniques 
is beyond the scope of this paper, 
but it is worth making the following remarks.

It is noted in \cite{Dessieux_2023a} that if $\chi=0$, 
parameters of sinusoids can be determined using the conventional Hough transform.
In this case, eq.(\ref{eq:for_g}) simplifies to 
$\lambda(\varphi) = d_1 k^{\tt{S}}_1(\varphi)+d_2 k^{\tt{S}}_2(\varphi) = d_1 \cos \varphi+d_2 \sin \varphi$.
With $x_i=x_i(\varphi,\lambda)= k^{\tt{S}}_i(\varphi)/\lambda$ ($i=1,2$),
this relationship takes 
the (intercept) form of the equation for a line: 
$d_1 x_1 + d_2 x_2 = 1$.
Thus, if the image in coordinates $(\varphi,\lambda)$ is transformed 
to coordinates $(x_1,x_2)$,
the curve described by the parametric equations $x_i(\varphi) = k^{\tt{S}}_i(\varphi)/\lambda(\varphi)$
is a line. 
In other words, 
sinusoids in the original image become straight lines in the  transformed image.
Since values of $\lambda$ are positive, the image space $(x_1,x_2)$ is bounded, so 
lines in the image can be detected using Hough transform.\footnote{With 
Duda-Hart parameters $(\rho, \theta)$ of a line 
and the expression
$\rho= x_1 \cos \theta +  x_2 \sin \theta$ usually used for performing Hough transform  \cite{Duda_1972},
the first two components of $\mathbf{d}$ are 
$d_1=\cos \theta/\rho$ and $d_2=\sin \theta/\rho$,
and the dependence of the wavelength on the angle $\varphi$ 
takes the form $\lambda(\varphi)=\cos(\theta-\varphi)/\rho$.}
The question arises about an analogous procedure in the case of $0 < \chi < \pi/2$.
With such  $\chi$, eq.(\ref{eq:for_g}) can be written as 
$d_1 x_1 + d_2 x_2   + d_3 x_3   = 1$, 
where $x_i=x_i(\varphi,\lambda)= k^{\tt{S}}_i(\varphi)/\lambda$ ($i=1,2,3$).
Thus, the curve described by the parametric equations $x_i(\varphi) = k^{\tt{S}}_i(\varphi)/\lambda(\varphi)$ 
in three-dimensional space
belongs to a plane.
Formally, the problem of determining $\mathbf{d}$ comes down 
to detecting planar features in three-dimensional image space,
and there are algorithms for that, e.g., \cite{Sarti_2002,Bauer_2008,Limberger_2015}.
However, with limited wavelength windows and diffraction patterns containing only parts of sinusoids, 
the planar curves in the three-dimensional image space are short and 
such detection is unlikely to work in practice.

Parameters of the sinusoids can be determined using 
other methods. 
There a numerous dedicated algorithms for detection of sinusoids of fixed period in digital images.
Typical sinusoid detection methods rely on 
Hough transform modified to find sinusoids rather than straight lines \cite{Thapa_1997,Glossop_1999}.
These searches are exhaustive but computationally expensive. 
Therefore, alternative approaches avoiding the Hough transformation have also been proposed;
see, e.g., \cite{Moran_2020,Dias_2020,Sattarzadeh_2022}.
Sinusoid detection can be difficult due to intersections and overlapping of the curves.
Moreover, only some of the algorithms for detection of sinusoids are applicable 
to images obtained from transmission spectra
because fragments of sinusoids are missing from the images
due to wavelength window. 

A simple way to determine $\mathbf{d}$ is to detect position of a dip in spectrum 
for a selected $\varphi_{j}$, then check the continuity of the sinusoid in the adjacent spectra 
at $\varphi_{j \pm 1}$, and use (\ref{eq:fin_d}) to get approximate $\mathbf{d}$.
Knowing the approximate $\mathbf{d}$, spectra at  $\varphi_{j \pm 2}$, $\varphi_{j \pm 3}$
et cetera are checked for dips 
to validate this $\mathbf{d}$ and improve its accuracy or discard it and try again.
With this approach, the limitation of changing the angle $\varphi$ in equiangular steps can be abandoned. 
One can use an adaptive selection of $\varphi_{j}$ depending on the results 
obtained in the previous steps. 
This may reduce the number of sample orientations needed
and 
shorten the overall measurement time.

\section{Indexing of data obtained with inclined rotation axis}

\subsection{Indexing for orientation determination}

By determining $\mathbf{d}$ and then
$\mathbf{g}$ from a sinusoid on an experimental pattern, one gets a single approximate relationship
$O \mathbf{g} \approx \mathbf{h}$ with unknown orientation $O$ and $\mathbf{h}$ 
corresponding to one of the plausible reflecting planes. 
As a reciprocal lattice vector, $\mathbf{h}$ is an integer combination of 
known basis vectors $\mathbf{a}^i_{\tt{C}}$ 
given in the Cartesian reference system attached to the crystal, 
i.e.,  $\mathbf{h} =\sum_{i=1}^3 h_i \mathbf{a}^i_{\tt{C}}$, 
where $(h_1 \, h_2 \, h_3) = (hkl)$ are integers. 
Since there are many, say $M$, potential reflecting planes, 
one has the same number of vectors $\mathbf{h}_m$ ($m=1,2,\dots,M$). 
Given multiple sinusoids, one has multiple, say $N$, vectors $\mathbf{g}_n$.
The goal is to get $O$ and to ascribe some of the vectors $\mathbf{h}_m$
to vectors  $\mathbf{g}_n$.
One can express the relationship between the sets of vectors $\mathbf{g}_n$ and  $\mathbf{h}_m$
as
\bEq
O G \approx  H P \  ,
\label{eq:OG_HP}
\eEq
where 
$G=[\mathbf{g}_1 \  \mathbf{g}_2 \ldots \mathbf{g}_N]$,
$H=[\mathbf{h}_1 \  \mathbf{h}_2  \ldots   \mathbf{h}_M]$
and 
$P$ is 
an unknown $M \times N$ matrix with zero entries everywhere except
the values of $1$ at entries $mn$ such that $\mathbf{h}_m$
corresponds to $\mathbf{g}_n$. 
The point is to determine $O$ and $P$. 
The matrix $O$ solves the orientation
determination problem,
and 
$P$  solves the index assignment 
problem.
This formulation is 
the same as in other cases of indexing for crystal orientation determination;
see \cite{Morawiec_2020,Morawiec_2022}.
Therefore, the problem can be solved automatically using any reasonably general software
for crystal orientation determination.

\subsection{Ab initio indexing}

Knowing a number of reciprocal lattice vectors $\mathbf{g}_n$, one can 
determine parameters of the crystal lattice. For each vector, one has
\bEq
\mathbf{g}_n \cdot \mathbf{a}_i^{\tt{S}} \approx h^n_i \ , 
\label{eq:ab_initio}
\eEq
where $(h^n_1 \, h^n_2 \, h^n_3)=( h^n \, k^n \, l^n )$ are unknown integers
and $\mathbf{a}_i^{\tt{S}}$ ($i=1,2,3$) are unknown basis vectors of the direct lattice 
in the sample reference frame. 
Solving the above problem (i.e., determining the vectors $\mathbf{a}_i^{\tt{S}}$ and the indices 
$h^n_i$) is the essence of ab initio indexing.
It can be performed using one of the programs
designed for diffraction data; see, e.g., \cite{Duisenberg_1992,Morawiec_2017,Morawiec_2022}. 
One needs to note that software for ab initio indexing 
usually proposes multiple solutions, and the choice of the ultimate solution is left to the user.

The related problem of refinement of lattice parameters also relies on eq.(\ref{eq:ab_initio}).  
The task is to determine the direct lattice basis vectors $\mathbf{a}_i^{\tt{S}}$ from
the known vectors $\mathbf{g}_n$ and their indices $h^n_i$ ($n=1,2,\ldots,N$).
Using the $3 \times N$  matrix $\tilde{H}$ with the integer entries $\tilde{H}_{in}=h^n_i$, 
eq.(\ref{eq:ab_initio}) can be written in the form 
$G^T [ \mathbf{a}_1^{\tt{S}} \, \mathbf{a}_2^{\tt{S}} \, \mathbf{a}_3^{\tt{S}}] \approx \tilde{H}^T$,
and one looks for 
the best matching matrix $[ \mathbf{a}_1^{\tt{S}} \, \mathbf{a}_2^{\tt{S}} \, \mathbf{a}_3^{\tt{S}}]$.
The least-squares solution to the problem can be written as 
$$
[ \mathbf{a}_1^{\tt{S}} \, \mathbf{a}_2^{\tt{S}} \, \mathbf{a}_3^{\tt{S}}] = 
(\tilde{H} G^{+} )^T \ .
$$
Such refinement is usually the final operation of ab intio indexing.
Needless to say, the above formula is also the basis for determining elastic strain tensors.

\subsection{Compliance check}

The proposed strategy of data analysis comes down to the following:
Based on the coordinates $(\varphi,\lambda)$ of points of a sinusoid, 
its parameters $\mathbf{d}$ are calculated using (\ref{eq:fin_d}), 
and then the corresponding reciprocal 
lattice vector $\mathbf{g}$ is obtained from~(\ref{eq:for_gs}).
This step is repeated for several sinusoids.
Thus obtained set of reciprocal lattice vectors $\mathbf{g}_n$ ($n=1,2,\ldots,N$) allows for determination 
of crystal orientation or lattice parameters using existing indexing software.
The practical usefulness of the 
strategy
will 
depend on the accuracy of the $\mathbf{g}_n$ vectors.
The lower the accuracy the more difficult the indexing. 
While indexing ideal data is straightforward, 
it is worth confirming that the scheme for 
determination of the $\mathbf{g}_n$ vectors is consistent with the indexing procedures.

\begin{figure}[t]
	\begin{picture}(300,190)(0,0)
		\put(50,0){\resizebox{12.0 cm}{!}{\includegraphics{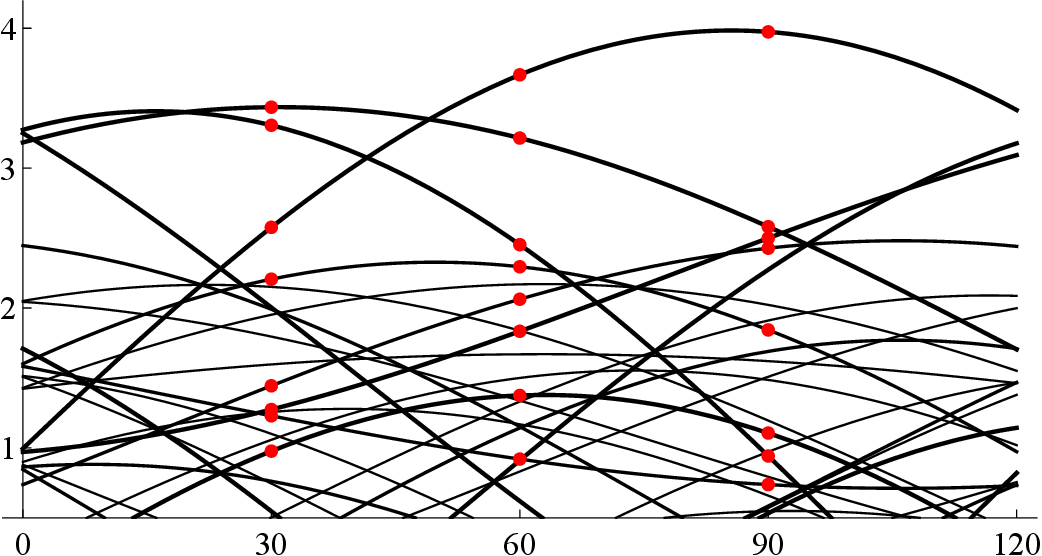}}}
		\put(29,182){$\lambda$ [\AA] }
		\put(397,1){$\varphi$ [deg] }
		\put(160,175){$\chi=35.264^{\circ}$ }
	\end{picture}
	\vskip 0.0cm
	\caption{Pattern in the wavelength window $0.5-4.2$\mbox{\AA} and 
		$\varphi$ in the range from $0$ to
		$120^{\circ}$. It is a part of the pattern shown in Fig.~\ref{Fig_completePatternS}\textit{a}.
		Red disks mark points of the eight sinusoids used for compliance check.
	}
	\label{Fig_windowS}
\end{figure}
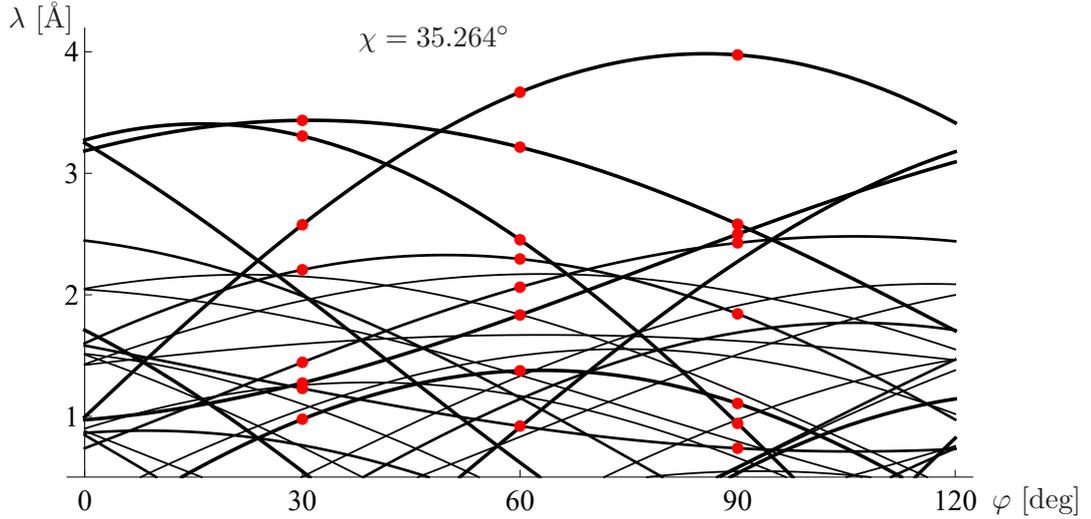

To illustrate the consistency test, data from the window shown in Fig.~\ref{Fig_windowS} are used.
Wavelengths of eight deepest dips 
in spectra for $\varphi=30^{\circ}$, $60^{\circ}$ and  $90^{\circ}$ were calculated.
Based on eqs.~(\ref{eq:system}), eight vectors $\mathbf{g}_n$ were obtained.
These vectors were input to the orientation determining program \textit{KiKoCh2} 
\cite{Morawiec_2020}. 
The resulting orientation was $(\overline{1}\,\overline{2}\,3)[6\,3\,4]$,
which is symmetrically equivalent to the original variant of the $S$ orientation. 
The same set of $\mathbf{g}_n$  vectors was used as input to the program \textit{Ind\_X} 
\cite{Morawiec_2017} to test ab initio indexing.
With suitably chosen parameters of \textit{Ind\_X}, the program provided  
a primitive lattice cell which, after  
application of \textit{LEPAGE} \cite{Spek_1988}, led to the conventional
face centered cubic cell with $a=3.613$\AA.

Summarizing,  in both cases (indexing for orientation determination and ab initio indexing), 
the data used to simulate the pattern were recovered, which means that 
the compliance tests confirmed the suitability of the proposed strategy.

\section{Concluding remarks}

Wavelength-resolved neutron imaging can be used for characterization of crystal lattices.
Attenuation of the transmitted beam caused by diffraction leads to Bragg dips in transmission spectra 
at positions corresponding to diffraction reflections. 
If spectra are collected at different sample orientations, the 
positions of the dips provide a basis for determination of crystal 
orientation and lattice parameters.
The simplest way to change sample orientation is by stepwise rotation 
about a fixed axis with spectrum recorded at each step.
The resulting pattern of intensities depends on the rotation angle and the wavelength
and contains characteristic sinusoids. 
Each sinusoid represents dependence of wavelength on the angle of rotation 
for reflections from a crystal plane. 
The sinusoid can be used for determination of coordinates of the reciprocal lattice vector
corresponding to the plane. 
Crystal orientations and lattice parameters are determined from sets of such vectors.

The usual approach is to rotate the sample about an axis perpendicular to the beam direction. 
However, it is shown above that the rotation about such an axis is not optimal for 
determination of the reciprocal lattice vectors from the sinusoids. 
A reciprocal lattice vector can be uniquely determined 
from coordinates of points of the corresponding sinusoid
only if the rotation axis is inclined to the beam direction. 
It is also shown that having a set of such vectors, one can compute crystal 
orientation or lattice parameters using existing indexing software.

The above considerations concern patterns composed 
of multiple spectra obtained from a single crystal, but the presented results are  
also relevant for indexing individual spectra, and also for 
orientation mappings of multigrain samples.
Moreover, 
the considerations are focused on the configuration with a fixed axis of sample rotation,
but they can be generalized to indexing of 
data obtained from spectra collected at arbitrarily set sample orientations.

\correction{The manuscript covers only theoretical aspects of 
	determination of lattice geometry from neutron transmission spectra
	and  ignores experimental issues. 
	It should be noted that on the experimental side, 
	the key to accurate reciprocal lattice vectors is the wavelength resolution.
	Neutron time-of-flight spectra are characterized 
	by a high density of peaks at short wavelengths 
	(i.e., at the bottom of patterns of sinusoidal curves), and
	sufficiently high resolution is needed to resolve the peaks
	and to get their positions.
}

\section*{Appendix: Indexing Bragg dips in single spectrum}

A simple approach to indexing of individual Bragg-dip patterns
is based on the use of cut-off wavelength equal to doubled interplanar spacing
of a family of reflecting planes \cite{Santisteban_2005,Shishido_2023}:
If a dip is at a wavelength larger than the cut-off wavelength
of a family, it must be due to reflection from a stack of
planes with larger spacing.
This helps in determining indices of the first dip at the largest wavelengths.
The remaining dips are indexed by considering orientations rotated about axis 
perpendicular to the stack of planes responsible for the first dip. 
Also forward indexing has been used for analyzing individual spectra 
\cite{Sato_2017}: Wavelengths of dips were computed 
for a grid over the domain of transmitted beam directions, 
and the beam direction was determined by matching 
the experimental Bragg-dip pattern to these simulated data.

A Bragg-dip pattern can be indexed by other 
methods suitable for automation.
The most robust ones are based on accumulation of contributions from 
some primitive configurations. 
Such methods are used for feature extraction  
from data (e.g., Hough transform) and also 
in indexing for orientation determination \cite{Ohba_1981,Morawiec_2022}.
A simple algorithm of this type 
(with pairs of Bragg dips voting for beam directions)
is sketched below. 
For this purpose, one needs a formula for computing the beam direction 
from positions of two Bragg dips 
and corresponding $\mathbf{h}$ vectors.

\vskip 0.2cm

\noindent
\textit{Beam direction from positions and indices of two Bragg dips}
\\
The direction of the incident beam in the crystal reference frame is 
$\widehat{\mathbf{k}}^{\tt{C}} =  O \widehat{\mathbf{k}}^{\tt{S}} =  O R_{\varphi}^T \widehat{\mathbf{k}}_0$,
and eq.(\ref{eq:intemediate}) can be written in the form
$ 
\mathbf{h} \cdot \widehat{\mathbf{k}}^{\tt{C}} + \lambda \, \mathbf{h}^2/2 = 0
$. 
Relationships
$$ 
\mathbf{h}_i \cdot \widehat{\mathbf{k}}^{\tt{C}} + \lambda_i \, \mathbf{h}_i^2/2 = 0 \ , \ \  i=1,2 
$$ 
between wavelengths $\lambda_1$, $\lambda_2$ and independent vectors $\mathbf{h}_1$,  $\mathbf{h}_2$ 
plus the normalization condition 
$\widehat{\mathbf{k}}^{\tt{C}}  \cdot \widehat{\mathbf{k}}^{\tt{C}} =1$
constitute a system of three equations for $\widehat{\mathbf{k}}^{\tt{C}}$.
The solution to this system is
\bEq
\widehat{\mathbf{k}}^{\tt{C}} = 
\frac{\mathbf{y} \times \mathbf{z} \pm \mathbf{y} \sqrt{4 \mathbf{y}^2-\mathbf{z}^2 }}{2 \mathbf{y}^2}
\ ,
\label{eq:kCvects}
\eEq
where
$\mathbf{y} =\mathbf{h}_1 \times \mathbf{h}_2$, 
$\mathbf{z} =\left( \lambda_1 \mathbf{h}_1^2 \right)  \mathbf{h}_2 - 
\left( \lambda_2 \mathbf{h}_2^2 \right)  \mathbf{h}_1$
and $4 \mathbf{y}^2  \geq \mathbf{z}^2$. 
Thus, if the dips at $\lambda_1$ and $\lambda_2$ are due to 
$\mathbf{h}_1$ and  $\mathbf{h}_2$, respectively, the beam direction is 
along one of the vectors given by eq.(\ref{eq:kCvects}).
Clearly, if the pairs
$\lambda_1$, $\mathbf{h}_1$ and  $\lambda_2$, $\mathbf{h}_2$ are exchanged, 
one obtains the same vectors $\widehat{\mathbf{k}}^{\tt{C}}$.
If vectors $\mathbf{h}_1$ and $\mathbf{h}_2$ are replaced by $-\mathbf{h}_1$ and $-\mathbf{h}_2$,
respectively, then $\mathbf{y}$ remains unchanged and $\mathbf{z}$ changes to $-\mathbf{z}$.
Hence, if the pair $\mathbf{h}_1$ and $\mathbf{h}_2$ leads to 
the pair of vectors $\widehat{\mathbf{k}}^{\tt{C}}$ given by eq.(\ref{eq:kCvects}),
 $-\mathbf{h}_1$ and $-\mathbf{h}_2$  lead to 
the pair of opposite vectors. 

\vskip 0.2cm

\noindent
\textit{Voting for beam direction}
\\
Indexing of Bragg dips in a single spectrum based on known crystal lattice
is equivalent to determination
of the transmitted beam direction in the crystal coordinate system. 
The algorithm for determination of the beam direction can be analogous to 
one of those used for orientation determination \cite{Morawiec_2022}.
One knows the positions of dips $\lambda_k$ ($k=1,2,\ldots,K$). 
Knowing lattice parameters and families of reflecting planes,
one has the reciprocal lattice vectors $\mathbf{h}_m$ ($m=1,2,\ldots,M$).
The 
domain of the beam direction depends on the crystal point symmetry.
(E.g., in the cubic case, it is the standard triangle.) 
The domain constitutes the parameter space.
In the simplest approach, the space is divided into bins. 
The first step is to set the counters assigned to the bins to zero. 
The main computations are based on
loops over all pairs of wavelengths $(\lambda_{k_1}, \lambda_{k_2})$,
and over  
all pairs of independent vectors $(\mathbf{h}_{m_1},\mathbf{h}_{m_2})$.
For given $(\lambda_{k_1}, \lambda_{k_2})$ and $(\mathbf{h}_{m_1},\mathbf{h}_{m_2})$,
using eq.(\ref{eq:kCvects}), 
one obtains a pair of $\widehat{\mathbf{k}}^{\tt{C}}$ vectors or no solution;
in the former case
the counters of the bins containing $\widehat{\mathbf{k}}^{\tt{C}}$ are increased by 1.
\correction{As a result of these computations, the counters assigned to the bins 
take on new values.}
The vector $\widehat{\pmb{\kappa}}^{\tt{C}}$ corresponding to the 
bin with the largest counter is the beam direction. 
In the end, it is worth refining $\widehat{\pmb{\kappa}}^{\tt{C}}$ by fitting dip positions.

Application of the above scheme to experimental dip positions listed in Table~1 of 
\cite{Santisteban_2005} led to 
$\widehat{\pmb{\kappa}}^{\tt{C}} = [0.7407 \ \,  0.6717 \ \, 0.0142 ]^T$ ($2.9^{\circ}$ away from 
$\langle 1 \, 1 \, 0 \rangle$ declared in \cite{Santisteban_2005}).
The root-mean-square deviation between the experimental and computed positions
was $2.4 \times 10^{-4}$\AA.

\vskip 0.2cm

\noindent
\textit{Crystal orientation from beam directions}
\\
The beam directions obtained by indexing individual Bragg-dip patterns 
can be used for further crystallographic analyses.
For instance,  
they can be a basis for orientation determination: 
With $J$ ($J \geq3$) distinct transmission spectra indexed, one has 
the same number of resulting beam directions $\widehat{\pmb{\kappa}}^{\tt{C}}_j$  ($j=1,2,\ldots,J$),
each reduced to the domain used in the indexing procedure. 
The direction $\widehat{\pmb{\kappa}}^{\tt{C}}_j$ is equivalent to $S \widehat{\pmb{\kappa}}^{\tt{C}}_j$, 
where the matrix $S$ represents a symmetry operation from the crystal point group. 
For each pair $\widehat{\pmb{\kappa}}^{\tt{C}}_j$ and $\widehat{\mathbf{k}}^{\tt{S}}(\varphi_j)$,
there exists an operation $S_j$ such that these vectors are related by 
$S_j \widehat{\pmb{\kappa}}^{\tt{C}}_j  \approx  O \widehat{\mathbf{k}}^{\tt{S}}(\varphi_j)$. 
First, one needs to determine the operations $S_j$ based on the fact that 
the vectors $\widehat{\mathbf{k}}^{\tt{C}}_j=S_j \widehat{\pmb{\kappa}}^{\tt{C}}_j$ consistent with 
$\widehat{\mathbf{k}}^{\tt{S}}(\varphi_j)$
satisfy the relationships
$
\widehat{\mathbf{k}}^{\tt{C}}_{j_1} \cdot \widehat{\mathbf{k}}^{\tt{C}}_{j_2} \approx
\widehat{\mathbf{k}}^{\tt{S}}(\varphi_{j_1}) \cdot \widehat{\mathbf{k}}^{\tt{S}}(\varphi_{j_2})
$
for $j_1$ and $j_2$ in $\{ 1,2,\ldots,J \}$.
Clearly, the result is not unambiguous: if vectors $\widehat{\mathbf{k}}^{\tt{C}}_j$ satisfy these relations, 
then the vectors $-\widehat{\mathbf{k}}^{\tt{C}}_j$ also satisfy them.
With $K_{\tt{C}}=[\widehat{\mathbf{k}}^{\tt{C}}_1 \  \widehat{\mathbf{k}}^{\tt{C}}_2 
\ldots \, \widehat{\mathbf{k}}^{\tt{C}}_J ]$, $K_{\tt{S}}$ defined  in section \ref{sec:incl} 
and $\widehat{\mathbf{k}}^{\tt{C}}_j \approx  O \widehat{\mathbf{k}}^{\tt{S}}(\varphi_j)$, 
one has $K_{\tt{C}} \approx O K_{\tt{S}}$.
Hence, if  $0< \chi < \pi/2$, the crystal orientation is 
$$
O \approx  {\cal SO}(K_{\tt{C}} K_{\tt{S}}^+) \ ,
$$ 
where ${\cal SO}(A)$ denotes 
the special orthogonal matrix nearest to the matrix $A$.\footnote{If the singular value decomposition of 
$A$ is $A = U \Sigma V^T$, then 
${\cal SO}(A) = U S V^T$ where $S=\mbox{diag}[+1,+1,\mbox{sign}(\det(A))]$; 
see, e.g., \cite{Morawiec_2022}.} 
The choice between 
$K_{\tt{C}}=[\widehat{\mathbf{k}}^{\tt{C}}_1 \  \widehat{\mathbf{k}}^{\tt{C}}_2 \ldots \, \widehat{\mathbf{k}}^{\tt{C}}_J ]$
and 
$K_{\tt{C}}=-[\widehat{\mathbf{k}}^{\tt{C}}_1 \  \widehat{\mathbf{k}}^{\tt{C}}_2 \ldots \, \widehat{\mathbf{k}}^{\tt{C}}_J ]$
is resolved by the condition $\det(K_{\tt{C}} K_{\tt{S}}^+)>0$.

With a slight modification, the above approach can be used to obtain orientations 
from data collected with the crystal rotation axis perpendicular to the beam direction ($\chi=0$).
The choice of the symmetry operations $S_j$ needed to get  
$\widehat{\mathbf{k}}^{\tt{C}}_j=S_j \widehat{\pmb{\kappa}}^{\tt{C}}_j$
is resolved by comparing the angles between 
$\widehat{\mathbf{k}}^{\tt{C}}_{j_1}$ and $\widehat{\mathbf{k}}^{\tt{C}}_{j_2}$ 
to  $\varphi_{j_1}-\varphi_{j_2}$ 
(as vectors $\widehat{\mathbf{k}}^{\tt{C}}_j$ consistent with 
$\widehat{\mathbf{k}}^{\tt{S}}(\varphi_j)$ satisfy the relationship
$\widehat{\mathbf{k}}^{\tt{C}}_{j_1} \cdot  \widehat{\mathbf{k}}^{\tt{C}}_{j_2} \approx \cos(\varphi_{j_1}-\varphi_{j_2})$).
Since the third components of the vectors $\widehat{\mathbf{k}}^{\tt{S}}(\varphi_j)$ are zero, 
the third column of $K_{\tt{S}}^+$ is zero, and in effect the third column of 
$K_{\tt{C}} K_{\tt{S}}^+$ is zero,
i.e., $K_{\tt{C}} K_{\tt{S}}^+$ has the form $[\mathbf{q}_1 \, \mathbf{q}_2 \, \mathbf{0}]$.
The first two columns of $K_{\tt{C}} K_{\tt{S}}^+$ approximate the first two columns of $O$.
Since the matrix $O$ is special orthogonal, its third column is the vector product of its
first and second columns. 
Thus, the crystal orientation is
$O \approx  {\cal SO}([\mathbf{q}_1 \ \mathbf{q}_2 \ \mathbf{q}_1 \times \mathbf{q}_2])$.
With  
$K_{\tt{C}}=-[\widehat{\mathbf{k}}^{\tt{C}}_1 \  \widehat{\mathbf{k}}^{\tt{C}}_2 \ldots \, \widehat{\mathbf{k}}^{\tt{C}}_J ]$,
one obtains the second possible orientation
${\cal SO}([-\mathbf{q}_1 \ -\mathbf{q}_2 \ \, \mathbf{q}_1 \times \mathbf{q}_2]) \approx OC$. 
This is another exposition of  
the two-way ambiguity in determining orientation 
when $\chi=0$.

\newpage

\bibliographystyle{unsrt}
\bibliography{Index_Bragg_dip_patts.bib}

\end{document}